\documentclass{article}
\usepackage{amsfonts}
\usepackage{amsmath}
\usepackage{amssymb}
\usepackage{graphicx}
\usepackage{array}
\usepackage{booktabs}
\usepackage{multirow}
\usepackage{makecell}
\usepackage{float}
\usepackage{booktabs}
\pdfoutput=1 

\usepackage{mymacros}
\usepackage{chngcntr}
\usepackage{verbatim}
\usepackage{amsthm}
\usepackage{graphics}
\usepackage{xcolor}
\usepackage{mathrsfs}
\usepackage{bbold}
\usepackage{cases}
\usepackage{epsfig}
\usepackage{epstopdf}
\usepackage{hyperref}
\usepackage{amsfonts}
\usepackage{hyperref}
\usepackage{jheppub} 

\usepackage[T1]{fontenc} 

\hypersetup{hypertex=true,
	colorlinks=true,
	linkcolor=blue,
	anchorcolor=blue,
	citecolor=blue}
\usepackage[numbers,sort&compress]{natbib}
\usepackage{threeparttable}

\title{Report on chaos bound outside Taub-NUT black holes}
\author[a]{Yucheng He \footnote{E-mail: heyucheng365@hotmail.com }}
\author[a]{Zeqiang Wang \footnote{E-mail: zeqwang@hotmail.com}}
\author[a]{Deyou Chen \footnote{E-mail: deyouchen@hotmail.com }}

\affiliation{$^{a}$School of Science, Xihua University, Chengdu 610039, China}

\abstract{Positions of a charged particle's equilibrium orbits and spatial regions where the chaos bound is violated are found through circular motions of the particle around charged Taub-NUT black holes. Lyapunov exponent is gotten by calculating eigenvalues of a Jacobian matrix in a phase space ($r, \pi_r$). When the particle's charge is fixed, the positions of the equilibrium orbits gradually move away from the event horizons with the increase of the angular momentum.The result shows that the bound is violated in the near-horizon regions and at a certain distance from the horizons when the charge and NUT parameter are fixed. The spatial regions increase with the increase of the NUT parameter's value.} 

\keywords{Chaos bound, spatial regions, Taub-NUT black holes}

\begin{document}
	\maketitle

\section{Introduction}

Motions of particles near black holes convey important information on background spacetimes. For example, null  geodesics effectively explore quasinormal modes for test fields, and these modes are related to both the internal information and surface quantization of black holes \cite{KZS1,KZS2,HOD1,OD1,GK1}. Tunneling behaviors of particles near the event horizons can reflect the temperature of black holes \cite{PW1}. The formation of black holes' shadows can be understood by studying motions of photons around black holes \cite{EHT1,EHT2,EHT3}. Research on these motions can bridge black holes physics and information theory.

Due to the nonlinearity of the Einstein's field equation, motions of particles may cause chaos. Chaos is a nonlinear phenomenon that is very sensitive to initial conditions. This sensitivity is represented by a Lyapunov exponent. A lot of work has been done on the chaos and exponent \cite{Pappalardi:2022zbn,CamposDelgado:2022cwu,Caputa:2022eye,Balasubramanian:2022tpr,Baskan:2022dys,Giataganas:2021ghs,Rosenhaus:2021xhm,Gong:2022nwp,Blake:2021hjj,Qu:2021ius,Pappalardi:2021ahe,Susskind:2021esx,Kundu:2021mex,Craps:2021bmz,Morita:2021syq,Chan:2020ujs,Nosaka:2020nuk,Lei:2020clg,Liu:2020yaf,Bianchi:2020des}. Focusing on the near-horizon region of the black hole, Hashimoto and Tanahashi studied the chaos generated by the radial motion of the particle \cite{HT}. The particle is subjected to an strong external force so that it can come very close to the hole without falling into it. They found the value of the exponent is independent on the strength and species of the external force and determined by the surface gravity of the hole,

\begin{equation}
\lambda \le \kappa,
\label{eq1.1}
\end{equation}

\noindent where $\lambda$ is the exponent and $\kappa$ is the surface gravity. This result favorably supported the conjecture proposed recently by Maldacena, Shenker and Stanford \cite{MSS}. In this seminal conjecture, they proposed that there is a general upper bound for the Lyapunov exponent of chaos in quantum field theory with a large number of degrees of freedom. The exponent satisfies

\begin{equation}
\lambda \le \frac{2\pi T}{\hbar},
\label{eq1.2}
\end{equation}

\noindent where $T$ is the system's temperature originally derived by thought experiments of shock waves near black holes' horizons \cite{JM}. For a black hole, its temperature is determined by the surface gravity. Therefore, Eqs. (\ref{eq1.1}) and (\ref{eq1.2}) are equivalent. This bound has been extensively studied and confirmed by a large amount of work. It was proved to be saturated in Sachdev-Ye-Kitaev model and speculated that this saturation is dual to an Einstein gravity \cite{AK1,Kitaev_2018,IH1}.

In the recent work, Zhao et al considered the particles' equilibrium in the near-horizon regions, and expanded the exponent at the event horizons \cite{ZLL}. When the contribution of sub-leading terms was considered, they found that the upper bound of the exponent (chaos bound) is violated. Cases of the chaos bound for the violations have also been found in \cite{KG1,KG2}. In \cite{KG1,KG2}, Kan et al found that the bound is violated when the influence of the particle's angular momentum on the exponent was taken into account. In their work, the particle's mass and charge are fixed at certain values and the exponent was gotten by the effective potential method. Also considering the contribution of the angular momentum, Lei and Ge et al obtained the expression of the exponent by the matrix method, and studied the bound through the circular motion of the particle around the black hole and the exponent's expansion at the event horizon \cite{LG2}. They found that the bound was violated both in the near-horizon regions and at a certain distance from the horizons of Reissner-Nordstr\"om and Reissner-Nordstr\"om anti-de Sitter black holes.

In this paper, we investigate the influence of the angular momentum of a charged particle around the charged Taub-NUT black holes on the Lyapunov exponent, and find spatial regions where the chaos bound is violated. Although the exponent is also obtained by the matrix method, we always fix the particle's charge as a constant in our calculation. This is different from the work of Ge et al., in which the charge is not fixed. Therefore, our calculation is a special case of their work. The NUT charge has both rotation-like and electromagnetic charge-like characteristics. Therefore, it is interesting to investigate its influence on the exponent and bound.

The rest is organized as follows. In the next section, we review thermodynamics of the charged Taub-NUT black holes and deriving the exponent by calculating the eigenvalue of the Jacobian matrix in the phase space ($r,\pi_r$). In Section \ref{section3}, we investigate the influence of the particle's angular momentum and NUT charge on the exponent, and find the spatial regions where the bound is violated. The last section is devoted to our conclusions.

\section{Circular motion of particles around Taub-NUT black holes}
	
In this section we investigate the circular motion of a charged particle around the charged Taub-NUT black hole to derive the Lyapunov exponent. As an anisotropic cosmological model, the Taub-NUT solution is characterized by a Misner string similar to a singularity on the axis. This black hole is a solution of the Einstein-Maxwell theory. Its Lagrangian is
\begin{equation}
	\begin{aligned}
		S=\frac{1}{16}\int d^{4}x\sqrt{-g}(R-2\Lambda-F_{\mu\nu}F^{\mu\nu}).
	\end{aligned}
	\tag{2.1}
	\label{eq2.1}
\end{equation}

\noindent From the Lagrangian, the solution of the black hole is given by \cite{JGM}

 \begin{equation}
     	\begin{aligned}
     		ds^2=-\frac{f(r)}{r^2+n^2}(dt+2ncos\theta d\phi ) ^2+\frac{r^2+n^2}{f(r)}dr^2+(r^2+n^2)(d\theta^2+sin^2\theta d\phi^2),
     	\end{aligned}
      \tag{2.2}
      \label{eq2.2}
 \end{equation}

\noindent with the electromagnetic potential

\begin{equation}
A_{\mu}= \frac{-Q_{0}r}{r^2+n^2} \left(dt+2ncos \theta d\phi\right), \label{2.3}
\tag{2.3}
\end{equation}

\noindent where $ f(r)=r^2-2Mr-n^2+Q_{0}^2$, $M, n$, $Q_{0}$ are the mass, NUT parameter, and electric parameter of the black hole, respectively. This black hole has two Misner string singularities located at $\theta=0$ and $\theta=\pi$. There are two roots from $f(r)=0$, which yields event and inner horizons,

\begin{equation}
	r_{\pm}=M\pm\sqrt{M^2+n^2-Q_{0}^2}. \label{2.4}
	\tag{2.4}
\end{equation}

\noindent The surface gravity is

      \begin{equation}
      \kappa =\frac{r_+ - M}{r_{+}^2+n^2}. 
      \label{2.5} \tag{2.5}
      \end{equation}

\noindent  The Hawking temperature and the entropy are \cite{Bordo:2019slw}
 \begin{equation}
	T=\frac{1}{4\pi r_{+}}\left(1-\frac{Q_{0}^2}{r_{+}^2+n^2}\right),\quad S=\pi\left(r_{+}^2+n^2\right). \label{2.6}
	\tag{2.6}
\end{equation}

\noindent The electric charge $Q$ is expressed by the electric parameter $Q_0$, which is 

\begin{equation}
      	Q=\frac{Q_{0}(r_{+}^2-n^2)}{r_{+}^2+n^2}.
      	\label{2.7}
      	\tag{2.7}
   \end{equation}

\noindent Its electric potential is $\Phi= \frac{Q_{0} r_{+}}{r_{+}^2+n^2}$. $N$ is the Misner charge associated with the NUT parameter and $\Psi$ is its conjugate quantity

 \begin{equation}
	N=-\frac{4\pi n^3}{r_{+}}\left(1-\frac{Q_0^2(n^2+3r_{+}^2)}{(r_{+}^2+n^2)^2}\right), \quad \Psi= \frac{1}{8\pi n}.
	\label{2.8}
	\tag{2.8}
\end{equation}

\noindent The above thermodynamic quantities obey the first law of thermodynamics

 \begin{equation}
	dM=TdS+\Phi dQ+ \Psi dN. \label{2.9}
	\tag{2.9}
\end{equation}

When a particle with mass $m$ and charge $q$ moves in a circular motion in the equatorial plane of the charged black hole, its Lagrangian is

      \begin{equation}
      	\mathcal{L} =\frac{1}{2}\left[-f\dot{t}^2+\frac{\dot{r}^2}{f}+(r^2+n^2)\dot{\phi}^2\right]-qA_{t}\dot{t},
      	\label{2.10}
      	\tag{2.10}
      \end{equation}

\noindent where $f=\frac{f(r)}{r^2+n^2}$, $\dot{x}^{\mu}=\frac{dx^{\mu}}{d\tau}$ and $\tau$ is proper time. Using the above equation and generalized momenta $\pi_{\mu} =\frac{\partial\mathcal{L}}{\partial\dot{x}}$, we get

      \begin{equation}
        \pi_{t}=-f\dot{t}-qA_{t}=-E,\quad  \pi_{r}=\frac{\dot{r}}{f}, \quad \pi_{\phi}=(r^2+n^2)\dot{\phi}=L.
       \label{2.11}
       \tag{2.11}
      \end{equation}

\noindent In the above equation, $E$ and $L$ represent the energy and angular momentum of the particle, respectively. The Hamiltonian is

      \begin{equation}
      	   H=\frac{-(\pi_{t}+qA_{t})^2+\pi_{r}^2f^2+\pi_{\phi}^2(r^2+n^2)^{-1}f}{2f}.
      	  \label{2.12}
      	  \tag{2.12}
      \end{equation}

\noindent From the Hamiltonian, the equation of motion of the particle can be obtained, which is

      \begin{equation}
      \begin{aligned}
      	&\dot{t} =\frac{\partial H}{\partial \pi_{t}}=-\frac{\pi_{t}+qA_{t}}{f},\quad \dot{\pi_{t}}=-\frac{\partial H}{\partial t}=0,\quad \dot{r}=\frac{\partial H}{\partial\pi_{r}}=\pi_{r}f, \\
      	&\dot{\pi_{r}}=-\frac{\partial H}{\partial r}=-\frac{1}{2}\left[\pi_{r}^{2}f^{'}-\frac{2qA_{t}^{'}(\pi_{t}+qA_{t})}{f}+\frac{(\pi_{t}+qA_{t})^2f^{'}}{f^{2}}-\pi_{\phi}^{2}((r^2+n^2)^{-1})^{'}\right], \\
      	&\dot{\phi}=\frac{\partial H}{\partial \pi_{\phi}}=\frac{\pi_{\phi}}{(r^2+n^2)},\quad \dot{\pi_{\phi}}=-\frac{\partial H}{\partial\phi}=0,    	
        \label{2.13}
      \end{aligned}	\tag{2.13}
      \end{equation}

\noindent where $"'"$ denotes a derivative with respect to $r$. In this paper, we use the matrix method to derive the exponent in the phase space $(r,\pi_r)$. Therefore, we need to get the radial coordinate and momentum at a coordinate time $t$ from the equations of motion Eq.(\ref{2.13}),

      \begin{equation}\tag{2.14}
      \begin{aligned}
      	&\frac{\mathrm{d}r}{\mathrm{d}t}=\frac{\dot{r}}{\dot{t}}=-\frac{\pi_{r}f^2}{\pi_{t}+qA_{t}}\\
      	&\frac{\mathrm{d}\pi_{r}}{\mathrm{d}t}=\frac{\dot{\pi_{r}}}{\dot{t}}=-qA_{t}^{'}+\frac{1}{2}\left[\frac{\pi_{r}f^2f^{'}}{\pi_{t}+qA_{t}}+\frac{\pi_{t}+qA_{t}f^{'}}{f}-\frac{\pi_{\phi}^{2}((r^2+n^2)^{-1})^{'}f}{\pi_{t}+qA_{t}}\right].   	
       \label{2.14}
      \end{aligned}
      \end{equation}

\noindent The four-velocity of a particle obeys a normalization condition $g_{\mu\nu}\dot{x^{\mu}} \dot{x^{\nu}} =\eta$,  where $\eta=0$ describes a case of a massless particle and $\eta=-1$ denotes a case of a massive particle. The particle is charged is in this paper and the normalization condition yields $\pi_{t}+qA_{t}=-\sqrt{f\left[1+\pi_{r}^{2}f+\pi_{\phi}^{2}(r^2+n^2)^{-1}\right]}$. We define $F_{1}=\frac{\mathrm{d}r}{\mathrm{d}t}$ and $ F_{2}=\frac{\mathrm{d}\pi_{r}}{\mathrm{d}t}$. Then Eq.(\ref{2.14}) is rewritten as

      \begin{equation}\tag{2.15}
      \begin{aligned}
      		&F_{1}=\frac{\pi_{r}f^2}{\sqrt{f\left(1+\pi_{r}^{2}f+\pi_{\phi}^{2}(r^2+n^2)^{-1}\right)}}, \\
      		&F_{2}=-qA_{t}^{'}+\frac{(2\pi_{r}^{2}f+1)f^{'}}{2\sqrt{f\left(1+\pi_{r}^{2}f+\pi_{\phi}^{2}(r^2+n^2)^{-1}\right)}}-\frac{\pi_{\phi}^{2}((r^2+n^2)f)^{'}}{2\sqrt{f\left(1+\pi_{r}^{2}f+\pi_{\phi}^{2}(r^2+n^2)^{-1}\right)}}.
      		\label{2.15}
      \end{aligned}
      \end{equation}
\noindent In the phase space, the elements of the matrix is defined by

      \begin{equation}
       K_{11}=\frac{\partial F_{1}}{\partial r},\quad K_{12}=\frac{\partial F_{1}}{\partial \pi_{r}},\quad K_{21}=\frac{\partial F_{2}}{\partial r},\quad K_{22}=\frac{\partial F_{2}}{\partial \pi_{r}}.
       \label{2.16}
       \tag{2.16}
      \end{equation}

\noindent For a particle whose equilibrium orbit is a circular should satisfies $\pi_{r}=\frac{d\pi_{r}}{dt}=0 $. The location of the equilibrium orbit is calculated from this constraint and Eq.(\ref{2.15}). By calculating the eigenvalues of the matrix, the exponent of the chaos of the charged particle in the orbit is obtained as follows

      \begin{equation}
      	  \lambda^{2}=\frac{1}{4}\Big[\frac{f^{'}+\pi_{\phi}^{2}((r^2+n^2)^{-1}f)^{'}}{1+\pi_{\phi}^{2}(r^2+n^2)^{-1}}\Big]^{2}-\frac{1}{2}f\frac{f^{''}+\pi_{\phi}^{2}((r^2+n^2)^{-1}f)^{''}}{1+\pi_{\phi}^{2}(r^2+n^2)^{-1}}-\frac{qA_{t}^{''}f^2}{\sqrt{f(1+\pi_{\phi}^{2}(r^2+n^2)^{-1})}}.
          \tag{2.17}\label{2.17}
      \end{equation}

\noindent When the Lyapunov exponent $\lambda>0$, the system of charged particle is unstable and chaos system. When the charge of the particle is fixed, we will calculate the Lyapunov exponent by different parameters of the black hole and particle. Clearly, the charge and angular momentum of particle affect the Lyapunov exponent. When the angular momentum is neglected, we get

\begin{equation}
	\lambda^2=\frac{1}{4}(f^{'})^2-\frac{1}{2}ff^{''}-qA_{t}^{''}f^\frac{3}{2}.
	\tag{2.18}\label{2.18}
\end{equation}

\noindent At the event horizon, $f(r)=0$, and (\ref{2.17}) is reduced to

\begin{equation}
\lambda^2=\frac{1}{4}(f^{'})^2 =\kappa^2.
\tag{2.19}
\label{2.19}
\end{equation}

\noindent Clearly, the exponent is saturated at the horizon. This result is consistent with that obtained in \cite{HT}.

 \section{Bound on Lyapunov exponent\label{section3}}
   \subsection{Lyapunov exponent in non-extremal charged Taub-NUT black holes}

When a charged particle moves around the charged Taub-NUT black hole, we can make the position of the particle's equilibrium orbit near the horizon by adjusting the charge-to-mass ratio and angular momentum of the particle. Different values of the Lyapunov exponent represent different motions. $\lambda^2>0$, $\lambda^2=0$ or $ \lambda^2<0$ denote unstable, critical or stable motions, respectively. When $\lambda^2>\kappa^2$, the bound of the exponent is violated. We investigate the influence of the angular momentum of the particle on the  exponent and find the angular momentum's range and spatial region where the exponent is violated. We set $M=1,$ and $q=15$. The position $r_{0}$ of the equilibrium orbit is listed in the following tables.

        	\begin{table}[H]	
        	\begin{threeparttable}
        		\centering
        		\begin{tabular}{ccccccccc}
        			\hline
        			\multirow{6}{*}{$r_{0}$} & L & 0 & 1 & 3 & 5 & 7 & 10 & 15  \\ \cline{2-9}
        			
        			& $Q_{0}$=0.50 & 2.27589&	2.28942&	2.37311&	2.47695& 2.57133& 2.68608 &	2.82149   \\ \cline{2-9}
        			& $Q_{0}$=0.80 &2.03071&	2.03672&	2.07901&	2.14326&	2.21223&	2.30866	&2.43907   \\ \cline{2-9}
        			& $Q_{0}$=0.95 & 1.87711&	1.88132&	1.91193&	1.96138&	2.01774&	2.10121&	2.22118   \\ \cline{2-9}
        			& $Q_{0}$=1.28 & 1.04148&	1.04239&	1.0513&	1.07936&	1.13341&	1.22565&	1.35941  \\ \hline
        		\end{tabular}
        		\begin{tablenotes}
        			\item Table 1. Positions of equilibrium orbits of the charged particle around the charged Taub-NUT black hole. For $n=0.80$, the event horizon is located at $r_{+}=2.1789$ when $Q_{0}=0.50$, at $r_{+}=2.0000$ when $Q_{0}=0.80$, at $r_{+}=1.8587$ when $Q_{0}=0.95$, and at $r_{+}=1.0400$ when $Q_{0}=1.28$.
        		\end{tablenotes}
        	\end{threeparttable}
        \end{table}

 \begin{table}[H]
 	\begin{threeparttable}
 		\centering
 		\begin{tabular}{ccccccccc}
 			\hline
 			\multirow{6}{*}{$r_{0}$}&L & 0 & 1 & 3 & 5 & 7 & 10 & 15  \\ \cline{2-9}
 			&$Q_{0}$=0.70 & 1.8343 & 1.83861&1.87001& 1.92062& 1.97786&	2.06121 &2.17736 \\ \cline{2-9}
 			&$Q_{0}$=0.90 &1.59727& 1.59933	&1.61509& 1.64327& 1.67912& 1.73848	&1.83387 \\ \cline{2-9}
 			&$Q_{0}$=1.00 & 1.40263&1.40386 &1.41348 &1.43164& 1.45648 &1.50154&1.58273  \\ \cline{2-9}
 			&$Q_{0}$=1.05  &1.24104	&1.24178& 1.24767& 1.25935& 1.27653& 1.31119&1.38354  \\ \hline
 		\end{tabular}
 		\begin{tablenotes}
 			\item Table 2.  Positions of equilibrium orbits of the charged particles around the charged Taub-NUT black hole. For $n=0.40$, the event horizon is located at $r_{+}=1.81854$ when $Q_{0}=0.70$,at $r_{+}=1.59161$ when $Q_{0}=0.90$, at $r_{+}=1.4000$ when $Q_{0}=1.00$, at $r_{+}=1.23976$ when $Q_{0}=1.05$,.
 		\end{tablenotes}
 	\end{threeparttable}
 \end{table}

       \begin{table}[!ht]
       	\begin{threeparttable}
       		\centering
       		\begin{tabular}{ccccccccc}
       			\hline
       			\multirow{6}{*}{$r_{0}$}&L & 0 & 1 & 3 & 5 & 7 & 10 & 15   \\ \cline{2-9}
       			&n=0.30 & 1.67851&	1.68111&1.70073&1.73484&1.77677&1.84338&1.94522  \\ \cline{2-9}
       			&n=0.60 & 1.86561&	1.86989&1.90101&1.95129&2.0084&	2.09226&2.21081  \\ \cline{2-9}
       			&n=0.80 & 2.03071& 2.03672& 2.07901&2.14326&2.21223 & 2.30866& 2.43907 \\ \cline{2-9}
       			&n=0.90 & 2.12257&	2.12958&2.17805&2.24951&2.32427&2.42668&2.56276 \\
       			\hline
       		\end{tabular}
       		\begin{tablenotes}
       			\item Table 3. Positions of equilibrium orbits of the charged particles around the charged Taub-NUT black hole. For $Q_0=0.80$, the event horizon is located at $r_{+}=1.6708$ when $n=0.30$,  at $r_{+}=1.8485$ when $n=0.60$, at $r_{+}=2.0000$ when $n=0.80$,and at $r_{+}=2.0816$ when $n=0.90$.
       		\end{tablenotes}
       	\end{threeparttable}	
       \end{table}

     \begin{table}[H]
     	\begin{threeparttable}
     		\centering
     		\begin{tabular}{ccccccccc}
     			\hline
     			 \multirow{6}{*}{$r_{0}$}&L & 0 & 1 & 3 & 5 & 7 & 10 & 15  \\ \cline{2-9}
     			&n=0.20 & 1.20074&1.20124   &1.20525   &	1.21332&1.22553	   &1.25147	&1.31129  \\ \cline{2-9}
     			&n=0.40 & 1.40263&1.40386 &1.41348 &	1.43164&  1.45648 &	1.50154&	1.58273  \\ \cline{2-9}
     			&n=0.60 & 1.60689&  1.60919 &1.62669  &	1.65752&  1.69612&	1.75899&	1.85861  \\ \cline{2-9}
     			&n=0.80 &1.81537&1.81913 & 1.84669 &	 1.89195&1.94446 &	 2.02361&	2.13964   \\ \hline
     		\end{tabular}
     		\begin{tablenotes}
     			\item Table 4. Positions of equilibrium orbits of the charged particles around the charged Taub-NUT black hole. For $Q_{0}=1.00$, the event horizon is located at $r_{+}=1.2000$ when $n=0.20$, at $r_{+}=1.4000$ when $n=0.40$, at $r_{+}=1.6000$ when $n=0.60$, and at $r_{+}=1.8000$ when $n=0.80$.
     		\end{tablenotes}
     	\end{threeparttable}
        \end{table}

In the above tables, when we fixed the NUT parameter and increased the angular momentum of the particle, the positions of the equilibrium orbits move away the horizon, and finally tend to a certain position. When the angular momentum is zero or small enough, the positions of the equilibrium orbits are closed to the horizon. In Table 1 and 2, when the angular momentum and NUT parameter were fixed, the positions of the equilibrium orbits moves closed to the horizon with the increase of the value of the electric parameter.

\begin{figure}[H]
        	\centering
        	\includegraphics{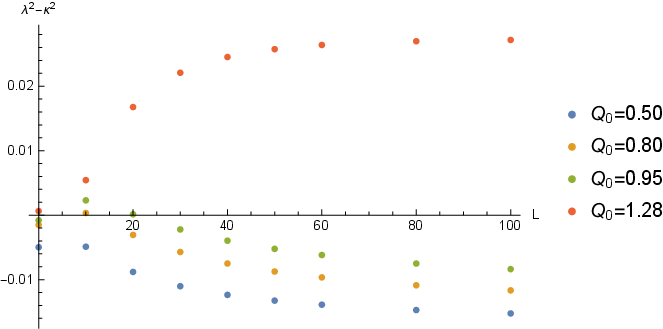}
        	\caption{The Lyapunov exponent of chaos of the charged particle outside the charged Taub-NUT black hole when $n=0.80$. The chaos bound is violated when $Q_{0}=0.80$ and $11.5558>L>4.8284$ (the corresponding spatial region is $2.4473>r_{0}>2.1356$), when $Q_{0}=0.95$ and $20.5396>L>2.4772$ ( $2.3247>r_{0}>1.9015$) and when $Q_{0}=1.28$ and $L>0$ ($2.2634>r_{0}>1.0414$).}
        	\label{fig1}
        \end{figure}

\begin{figure}[H]
	\centering
	\includegraphics{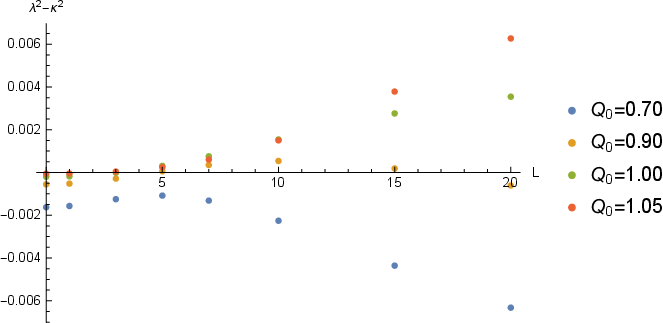}
	\caption{The Lyapunov exponent of chaos of the charged particle outside the charged Taub-NUT black hole when $n=0.40$.  The chaos bound is violated when $Q_{0}=0.90$ and $16.3314>L>4.6857$ (the corresponding spatial region is $1.8569>r_{0}>1.6382$), when $Q_{0}=1$ and $339.7641>L>3.1137,$ ($2.2391>r_{0}>1.4143$) and when $Q_{0}=1.05$ and $L>2.30113$ ($2.17152>r_{0}>1.24494$). }
	\label{fig3}
\end{figure}

\begin{figure}[H]
    	  \centering
    	  \includegraphics{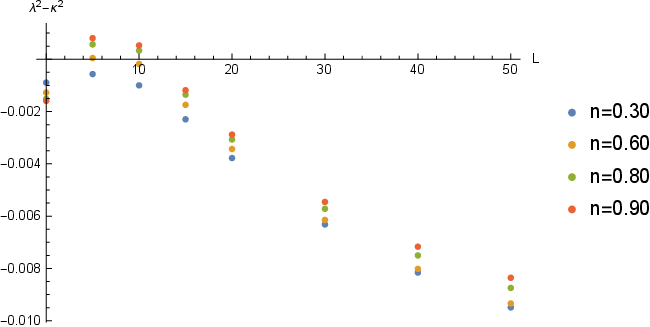}
    	  \caption{The Lyapunov exponent of chaos of the charged particle outside the charged Taub-NUT black hole when when $Q_{0}=0.80$. The chaos bound is violated when $n=0.60$, at $9.1781>L>4.7954,$ (the corresponding spatial region is $2.0700>r_{0}>1.9456$), when $n=0.80$, at $11.1126>L>3.5954,$ ($2.3412>r_{0}>2.0968$), when $n=0.90$, at $11.6743>L>3.2788,$ ($2.4769>r_{0}>2.1816$). }
    	  \label{fig2}
        \end{figure}

\begin{figure}[H]
        	\centering
        	\includegraphics{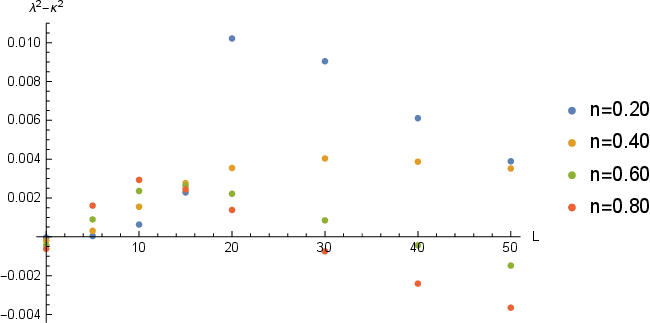}
        	\caption{The Lyapunov exponent of chaos of the charged particle outside the charged Taub-NUT black hole when $Q_{0}=1.00$. The chaos bound is violated in the range $L>3.96016$ ( $2.0895>r_{0}>1.2086$) when $n=0.20$, in the range $339.7641>L>3.1137$ ($2.2391>r_{0}>1.4143$) when $n=0.40$, in the range $36.3545>L>2.5581,$ ($2.1277>r_{0}>1.6215$) when $n=0.60$, and in the range $26.2532>L>2.1749$ ($2.3228>r_{0}>1.8325$) when $n=0.80$. }
        	\label{fig4}
        \end{figure}

Using Eq. \eqref{2.17}, we numerically calculate the values of the Lyapunov exponent at the equilibrium orbits and plot them in Figure \ref{fig1} - Figure \ref{fig4}. In Figure \ref{fig1}, when the parameter $Q_{0}$ is small, the value of the exponent increases to a maximum value with the increase of the angular momentum, and then decreases to a constant value. However, for any values of the angular momentum, the chaos bound is violated when the parameter $Q_{0}$ is large enough. In this case, the exponent does not have a maximum value and tends to a certain value with the increase of the angular momentum. The range of the angular momentum where the bound is violated are increased with the increase of the parameter $Q_{0}$. For different values of $Q_{0}$, the locations of equilibrium orbits and range of the angular momentum for the violation are different. The relative size of the spatial region of the particle's motion when the bound is violated is $1.22368>\frac{r_0}{r_+}>1.06782$ when $Q_{0}=0.80$, is $1.250729>\frac{r_0}{r_+}> 1.0230483$ when $Q_{0}=0.95$, and is $2.1763>\frac{r_0}{r_+}>1.0014$  when $Q_{0}=1.28$. Therefore, the relative size increases with the increase of the electric parameter's value.

When the NUT parameter is fixed at $n=0.40$, we calculate values of the  exponent at the equilibrium orbits with different values of the electric parameter. From Figure \ref{fig3}, we find that the violation still occurs when the electric parameter is relative large. The range of the angular momentum and spatial region for the violation increase with the increase of the electric parameter. The relative size of the spatial region is given by  $1.1626>\frac{r_0}{r_+}>1.0256$ when $Q_{0}=0.90$, by $1.5994>\frac{r_0}{r_+}>1.0316$ when $Q_{0}=1.00$ and by $1.75156>\frac{r_0}{r_+}>1.2449$ when $Q_{0}=1.05$.

When the electric parameter is fixed at $Q_{0}=0.80$, we calculate values of the exponent at the equilibrium orbits and plot them in Figure \ref{fig2}. In the figure, we find that a smaller value of the NUT parameter does not cause the violation for the bound, and a larger value of the parameter causes the violation When the angular momentum is within a certain range. When values of the NUT parameter are different, the equilibrium orbital positions and the range of the angular momentum are different. The relative spatial region is $1.1095>\frac{r_0}{r_+}>1.0429$ when $n=0.60$, is $1.17062>r>1.04843$ when $n=0.80$, and is $1.0995>\frac{r_0}{r_+}>1.0278$ when $n=0.90$.  From the figure, we find that the angular momentum's range and spatial region are increase with the increase of the NUT parameter's value. As this parameter increases, the black hole gradually approaches an extremal black hole. We infer that the extremal black hole is more likely to violate the bound. This will be discussed in the next section.

When the electric parameter is fixed at $Q_{0}=1.00$, we calculate values of the exponent at the equilibrium orbits and plot them in Figure \ref{fig4}. When the bound is violated, the minimum value and region of the angular momentum decreases as the NUT parameter's value increases. When the NUT parameter's values are different, the angular momentum range and relative spatial region are different. The relative spatial region is given by $1.7401>\frac{r_0}{r_+}>1.0065$ when  $n=0.20$, by $1.5994>\frac{r_0}{r_+}>1.0316$  when $n=0.40$, by $1.3298>\frac{r_0}{r_+}>1.0134$ when $n=0.60$, and is by $1.2904>\frac{r_0}{r_+}>1.0180$  when $n=0.80$. Clearly, the relative spatial region decreases with the increase of the NUT parameter's value.

For the non-extremal charged Taub-NUT black hole, the violation for the chaos bound are affected by the electric parameter, NUT parameter and the particle's angular momentum.

\subsection{Lyapunov exponent in extremal charged Taub-NUT black holes}

\begin{figure}[H]
     	\centering
     	\includegraphics{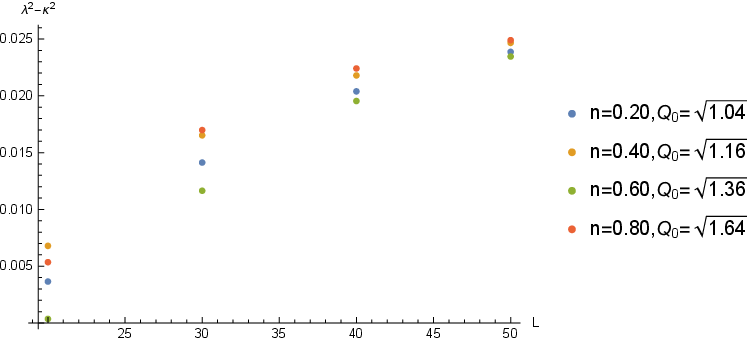}
     	\caption{The Lyapunov exponent of chaos of the charged particle outside the extremal Taub-NUT black hole when when. The chaos bound is violated in the range $L>14.0836$ ($2.0197>r_{0}>1.0000$) when $n=0.20$ and $Q_{0}=\sqrt{1.04}$, in the range $L>11.6518$ ($2.0771>r_{0}>1.0000$) when $n=0.40$ and $Q_{0}=\sqrt{1.16}$, in the range $L>8.1515$ ($2.16616>r_{0}>1.0000$) when $n=0.60$ and $Q_{0}=\sqrt{1.36}$, and in the range $L>4.0176$ ($2.28059>r_{0}>1.0000$) when $n=0.80$ and $Q_{0}=\sqrt{1.64}$.}
     	\label{fig5}
     \end{figure}

For an extremal charged Taub-NUT black hole, the inner and event horizons coincides with each other and the surface gravity is zero. There is $M^2+n^2=Q_{0}^2$ and $r_{+}=1$. Using \eqref{2.13}, we get positions of equilibrium orbits and list them  in Table 5. The relationship between the  Lyapunov exponent and angular momentum is plotted in Figure \ref{fig5}.

         \begin{table}[H]
         	\begin{threeparttable}
         		\centering
         		\begin{tabular}{cccccccc}
         			\hline
         			\multirow{6}{*}{$r_{0}$}&L & 0 & 10 & 20 & 30 & 40 & 50  \\ \cline{2-8}
         			&n=0.2,$Q_{0}=\sqrt{1.04}$  & * & * & 1.16866 & 1.35506 & 1.47461 & 1.55774  \\ \cline{2-8}
         			&n=0.4,$Q_{0}=\sqrt{1.16}$  & * & * & 1.23725 & 1.41428 & 1.53009 & 1.61158  \\ \cline{2-8}
         			&n=0.6,$Q_{0}=\sqrt{1.36}$  & * & 1.06827 & 1.33819 & 1.50427 & 1.6155 & 1.69495  \\ \cline{2-8}
         			&n=0.8,$Q_{0}=\sqrt{1.64}$  & * & 1.21692 & 1.46144 & 1.61723 & 1.72397 & 1.80138 \\ \hline
         		\end{tabular}
         		\begin{tablenotes}
         			\item Table5: Position of equilibrium orbits of a charged particle around the charged extremal Taub-NUT black hole. An asterisk indicates that a equilibrium orbit does not exist.
         		\end{tablenotes}
         		\label{table5}
         	\end{threeparttable}
         \end{table}

For the extremal case, when the particle's angular momentum is small, the positions of the equilibrium orbits does not exist and the particle drops into the black hole. A significant difference between it and non-extremal black hole is that the equilibrium orbit here can approach the event horizon infinitely without falling into it, which does not occur in the non-extremal black hole. The exponent is always positive at equilibrium orbits in the figure, which shows there always exists the violation for the chaos in the near-horizon region and at a certain distance from the horizon of the black hole. The range of the angular momentum and spatial region increase with the increase of the NUT parameter's value. Therefore, the extremal black hole is more likely to violate the chaos bound than the non-extremal black hole.

 \section{Conclusions}

In this paper, we investigated the influences of the particle's angular momentum and NUT parameter on the Lyapunov exponent and found the spatial regions where the chaos bound is violated by changing the angular momentum and fixing other parameters. The Lyapunov exponent was gotten by calculating the eigenvalues of the Jacobian matrix. For the non-extremal black hole, when the particle's angular momentum is fixed and the electric parameter increases, the equilibrium orbits are closer to the event horizon. With the increase of the NUT parameter, the spatial region for the violation is closer to the event horizon. Therefore, the relatively large electric parameter more likely violate the bound in the near-horizon region. For the extremal case, when the angular momentum is small, the positions of the equilibrium orbits do not exist and there always exists the violation. The extremal black hole is more likely to violate the chaos bound than the non-extremal black hole.

There are two explanations for the violation of the chaos bound. In \cite{LG2}, Ge et al believed that this violation was related to the stability of the black holes, and it is possible to solve this violation phenomenon through research on stability. In \cite{LWT1,LWT2}, the authors found the violation by exploring the influence of the minimum length effects on the chaotic motion. They believed that this result is not a violation for the conjecture proposed in \cite{MSS}, and the bound can be corrected by the minimum length in the bulk. The weak gravity conjecture reveals that the charge-to-mass ratio of a particle can be greater than 1. In this paper, we set the particle charge to 15, therefore, this conjecture is an implicit condition in our investigation. If this conjecture was not considered, the result maybe changed. In addition, the backreaction of the particle on the background spacetime has not been taken into account \cite{LG2}. It will be meaningful to study of the chaos bound by considering this effect.


\begin{thebibliography}{99}
	
\bibitem{KZS1}
R.A. Konoplya  and Z. Stuchlík, \emph{Are eikonal quasinormal modes linked to the unstable circular null geodesics?} \emph{Phys. Lett.B} \textbf{771} (2017) 597.

\bibitem{KZS2}
R.A. Konoplya, A.F. Zinhailo and Z. Stuchlík, \emph{Quasinormal modes, scattering and Hawking radiation in the vicinity of Einstein-dilaton-Gauss-Bonnet black hole}, \emph{Phys. Rev.D} \textbf{99} (2019) 124042.


\bibitem{HOD1}
S. Hod, \emph{Bohr’s correspondence principle and the area spectrum of quantum black holes}, \emph{Phys. Rev. Lett.} \textbf{81} (1998) 4293.
\bibitem{OD1}
O. Dreyer, \emph{Quasinormal modes, the area spectrum, and black hole entropy}, \emph{Phys. Rev. Lett.} \textbf{90} (2003) 081301.
\bibitem{GK1}
G. Kunstatter, \emph{D-dimensional black hole entropy spectrum from quasi-normal modes},\emph{ Phys. Rev. Lett.} \textbf{90} (2003) 161301.

\bibitem{PW1}
M.K. Parikh, and F. Wilczek,\emph{Hawking radiation ss tunneling}, \emph{Phys. Rev. Lett.} \textbf{85} (2000) 5024.

\bibitem{EHT1}
The Event Horizon Telescope Collaboration et al, \emph{First M87 event
horizon telescope results. I. The shadow of the supermassive black
hole}, \emph{Astrophys. J. Lett.} \textbf{875} (2019) L1.
\bibitem{EHT2}
The Event Horizon Telescope Collaboration et al, \emph{First M87 event
horizon telescope results. IV. Imaging the central supermassive
black hole}, \emph{Astrophys. J. Lett.} \textbf{875} (2019) L4.
\bibitem{EHT3}
The Event Horizon Telescope Collaboration et al, \emph{First M87 event
horizon telescope results. V. Physical origin of the asymmetric ring},
\emph{Astrophys. J. Lett.} \textbf{875} (2019) L5.

\bibitem{Pappalardi:2022zbn}
S.~Pappalardi and J.~Kurchan,
\emph{Quantum bounds on the generalized Lyapunov exponents},
\emph{Entropy} \textbf{25}, (2023) 246.

\bibitem{CamposDelgado:2022cwu}
R.~Campos Delgado and S.~Forste,
\emph{Lyapunov exponents in N=2 supersymmetric Jackiw-Teitelboim gravity},
\emph{Phys. Lett. B} \textbf{835}, (2022) 137550.



\bibitem{Caputa:2022eye}
P.~Caputa and S.~Liu,
\emph{Quantum complexity and topological phases of matter},
\emph{Phys. Rev. B} \textbf{106} (2022) 195125.

\bibitem{Baskan:2022dys}
K.~Ba\c{s}kan, S.~K\"urk\c{c}\"uo\u{g}lu and C.~Ta\c{s}c\i{},
\emph{Chaotic dynamics of the mass deformed ABJM model},
\emph{Phys. Rev. D} \textbf{107} (2023) 066006.

\bibitem{Balasubramanian:2022tpr}
V.~Balasubramanian, P.~Caputa, J.~M.~Magan and Q.~Wu,
\emph{Quantum chaos and the complexity of spread of states},
\emph{Phys. Rev. D} \textbf{106}  (2022) 046007.

\bibitem{Gong:2022nwp}
Z.~Gong and R.~Hamazaki,
\emph{Bounds in nonequilibrium quantum dynamics},
\emph{Int. J. Mod.} Phys. B \textbf{36}  (2022) 2230007.

\bibitem{Rosenhaus:2021xhm}
V.~Rosenhaus,
\emph{Chaos in a many-string scattering amplitude},
\emph{Phys. Rev. Lett.} \textbf{129}  (2022) 031601.

\bibitem{Giataganas:2021ghs}
D.~Giataganas,
\emph{Chaotic motion near black hole and cosmological horizons},
\emph{Fortsch. Phys.} \textbf{70} (2022) 2200001.

\bibitem{Blake:2021hjj}
M.~Blake and R.~A.~Davison,
\emph{Chaos and pole-skipping in rotating black holes},
\emph{JHEP} \textbf{2201} (2022) 013.

\bibitem{Qu:2021ius}
L.~C.~Qu, J.~Chen and Y.~X.~Liu,
\emph{Chaos and complexity for inverted harmonic oscillators},
\emph{Phys. Rev. D} \textbf{105}  (2022) 126015.

\bibitem{Pappalardi:2021ahe}
S.~Pappalardi, L.~Foini and J.~Kurchan,
\emph{Quantum bounds and fluctuation-dissipation relations},
\emph{SciPost Phys.} \textbf{12}  (2022) 130.

\bibitem{Susskind:2021esx}
L.~Susskind,
\emph{Entanglement and chaos in de Sitter space holography: an SYK example},
\emph{JHAP} \textbf{1}  (2021) 1.

\bibitem{Kundu:2021mex}
S.~Kundu,
\emph{Extremal chaos},
\emph{JHEP} \textbf{2201}  (2022) 163.

\bibitem{Craps:2021bmz}
B.~Craps, S.~Khetrapal and C.~Rabideau,
\emph{Chaos in CFT dual to rotating BTZ},
\emph{JHEP} \textbf{2111}  (2021)105.

\bibitem{Morita:2021syq}
T.~Morita,
\emph{Extracting classical Lyapunov exponent from one-dimensional quantum mechanics},
\emph{Phys. Rev. D} \textbf{106}  (2022) 10.

\bibitem{Chan:2020ujs}
A.~Chan, A.~De Luca and J.~T.~Chalker,
\emph{Spectral Lyapunov exponents in chaotic and localized many-body quantum systems},
\emph{Phys. Rev. Res.} \textbf{3} (2021) 023118.

\bibitem{Nosaka:2020nuk}
T.~Nosaka and T.~Numasawa,
\emph{Chaos exponents of SYK traversable wormholes},
\emph{JHEP} \textbf{2102} (2021) 150.

\bibitem{Lei:2020clg}
Y.~Q.~Lei, X.~H.~Ge and C.~Ran,
\emph{Chaos of particle motion near a black hole with quasitopological electromagnetism},
\emph{Phys. Rev. D} \textbf{104} (2021) 046020.

\bibitem{Liu:2020yaf}
Y.~Liu and A.~Raju,
\emph{Quantum chaos in topologically massive gravity},
\emph{JHEP} \textbf{2012} (2020) 027.

\bibitem{Bianchi:2020des}
M.~Bianchi, A.~Grillo and J.~F.~Morales,
\emph{Chaos at the rim of black hole and fuzzball shadows},
\emph{JHEP} \textbf{2005} (2020) 078.

\bibitem{HT}
K. Hashimoto and N. Tanahashi, \emph{Universality in chaos of particle motion near black hole horizon}, \emph{Phys. Rev.D} \textbf{95} (2017) 024007.

\bibitem{MSS}
J. Maldacena, S.H. Shenker and D. Stanford, \emph{A bound on chaos}, \emph{JHEP} \textbf{1608} (2016) 106.

\bibitem{JM}
J. Maldacena,
\newblock \emph{The large-N limit of superconformal field theories and supergravity},
\newblock {\em Int. J. Theor. Phys.} \textbf{38} (1999) 1113.

\bibitem{Kitaev_2018}
A. Kitaev and S.~Josephine Suh,
\newblock \emph{The soft mode in the sachdev-ye-kitaev model and its gravity dual},
\newblock {\em JHEP} \textbf{1805} (2018) 183.

\bibitem{AK1}	
A. Kitaev, \emph{Hidden Correlations in the Hawking Radiation and Thermal Noise},\emph{ talk
given at Fundamental Physics Prize Symposium, Nov. 10, 2014.
Stanford SITP seminars, Nov. 11 and Dec. 18, 2014.}
\bibitem{IH1}
I. Heemskerk, J. Penedones, J. Polchinski and J. Sully, \emph{Holography from Conformal
Field Theory}, \emph{JHEP} \textbf{0910} (2009) 079.

\bibitem{ZLL}
Q.Q. Zhao, Y.Z. Li and H. L$\ddot{u}$, \emph{Static equilibria of charged particles around charged black holes: Chaos bound and its violations}, \emph{Phys. Rev.D} \textbf{98} (2018) 124001.

\bibitem{KG1}
N. Kan and B. Gwak, \emph{Bound on the Lyapunov exponent in Kerr-Newman black holes via a charged particle}, \emph{Phys. Rev.D} \textbf{105} (2022) 026006.

\bibitem{KG2}
B. Gwak, N. Kan, B.H. Lee and H. Lee, \emph{Violation of bound on chaos for charged probe in Kerr-Newman-AdS black hole}, \emph{JHEP} \textbf{2209} (2022) 026.

\bibitem{LG2}
Y.Q. Lei and X.H. Ge, \emph{Circular motion of charged particles near a charged black hole}, \emph{Phys. Rev.D} \textbf{105} (2022) 084011.

\bibitem{JGM}
J. G. Miller, \emph{Global analysis of the Kerr‐Taub‐NUT metric}, \emph{J. Math. Phys.} \textbf{14} (1973) 486.

\bibitem{Bordo:2019slw}
A.~B.~Bordo, F.~Gray and D.~Kubiz\v{n}\'ak,
\emph{Thermodynamics and Phase Transitions of NUTty Dyons},
\emph{JHEP} \textbf{07} (2019) 119.


\bibitem{LWT1}
F. Lu, J. Tao and P. Wang,
\newblock \emph{Minimal length effects on chaotic motion of particles around black
hole horizon},
\newblock {\em JCAP} \textbf{1812} (2018) 036.

\bibitem{LWT2}
X. B. Guo, K. K. Liang, B. R. Mu, P. Wang and M. T. Yang,
\newblock \emph{Minimal length effects on motion of a particle in rindler space,} \emph{Chin. Phys. C}  \textbf{45} (2021) 023115.












	
	
	
	
	
	

	

	

	
\end{thebibliography}
\end{document}